\documentclass[aps,preprint,groupedaddress]{revtex4}

\usepackage{graphicx}%

\begin{document}

%\begin{frontmatter}

% You should use BibTeX and apsrev.bst for references
%\bibliographystyle{elsarticle-num}
\bibliographystyle{apsrev4-2}


\title{Solving the Crystallographic Phase Problem using Dynamical Scattering in Electron Diffraction}

\author{Christoph T. Koch}
%\email[]{christoph.koch@hu-berlin.de}

\affiliation{Department of Physics, Humboldt-Universit\"at zu Berlin, Berlin, Germany}

\date{\today}

\begin{abstract} 

Solving crystal structures from kinematical X-ray or electron diffraction patterns of single crystals requires many more diffracted beams to be recorded than there are atoms in the structure, since the phases of the structure factors can only be retrieved from such data if the atoms can be resolved as sharply peaked objects.
Here a method is presented by which the fact that multiple scattering encodes structure factor phases in the diffracted intensities is being used for solving the crystallographic phase problem. The retrieval of 
both amplitudes and phases of electron structure factors from diffraction patterns recorded with varying 
angle of incidence will be demonstrated. No assumption about the scattering potential 
itself is being made.  In particular, the resolution in the diffraction data does not 
need to be sufficient to resolve atoms, making this method potentially 
interesting for electron crystallography of 2-dimensional protein crystals 
and other beam-sensitive complex structures. 
%\footnote{This work is dedicated to the memory of John C.H. Spence.}.
% 
\end{abstract}

%\item Variation of diffracted intensity with beam tilt is used to solve the crystallographic phase problem.
%\item The crystallographic phase problem is solved independent of the number of diffracted beams
%\item Atomicity is not assumed, making this method applicable at any resolution. 

%\begin{keyword}
%electron diffraction \sep inversion of dynamical scattering \sep crystallography
%\PACS 61.05.jd \sep 61.05.jm \sep 61.05.cc
%\end{keyword}
%\end{frontmatter}

\maketitle

\section{Introduction}

Electron diffraction becomes the method 
of choice when trying to determine the atomic structure of small 
crystalline volumes
for mainly 3 reasons: a) electrons can be focused into very small probes matching the size of even the smallest volumes of material, b) their scattering strength is several orders of 
magnitude greater than that of X-rays, and c) in contrast to real-space based techniques, such as direct imaging or ptychography, the signal-to-noise ratio of the data increases linearly with (as opposed to the square root of) the number of unit cells being illuminated coherently.  For structure 
determination of inorganic materials the strong
interaction of electrons with the scattering medium is traditionally 
being viewed as a nuisance because existing methods for solving the phase 
problem in X-ray crystallography assume kinematic scattering, 
% i.e.\@ 
% that each scattered electron has scattered only once on its path through 
% the sample under investigation.  
making these methods generally not directly applicable to (single orientation) electron 
diffraction intensities. 
But even when reducing the discrepancy of electron diffraction data with a kinematic model, e.g.\@ by averaging over many different dynamical diffraction conditions as is done in precession electron diffraction \cite{Vincent94}, the shape of the crystal may not always permit the collection of diffraction data at a sufficiently high specimen tilt to ensure the high degree of completeness required by kinematic phasing techniques for resulting in a unique structure. 

% Solving the structure of highly beam-sensitive materials, such as 2-dimensional  protein crystals, from electron diffraction data alone, on the other hand, is hampered by that fact that these beam-sensitive samples do not  generally provide data of sufficient resolution to satisfy the preconditions for kinematic phasing methods.

% In this letter I will present a method by which the crystal 
% potential can be determined directly from dynamical electron diffraction 
% data by making use of the sensitivity of 
% dynamical diffraction intensities to structure factor \(n\)-phase 
% invariants (\(n \geq 3\)). 
% No assumption about the potential distribution 
% within the unit cell is being made.
% In particular, this method does not require the
% crystal potential to be a superposition of (sharply peaked) 
% atomic potentials, making it
% also applicable to very complex and/or beam-sensitive structures, such 
% as 2-dimensional protein crystals.

In contrast to kinematical diffraction intensities, which are absolutely insensitive to the phases of the structure factors, dynamical diffraction intensities, as observed in most electron diffraction patterns, are sensitive to the sums of phases of structure factor triplets and higher order multiplets, i.e.\@ sets of structure factors for which the sum of the corresponding reciprocal lattice vectors vanishes \cite{Moodie72}. While this encoding of structure factor phase information in the diffracted intensities is the basis for the refinement of structure factors, e.g.\@ from CBED patterns \cite{Zuo91,Zuo99}, it has inspired also a number of attempts to solve the 'dynamic inversion problem', i.e.\@ to retrieve structure factor amplitudes and phases from dynamical diffraction intensities alone and deterministically, i.e.\@ without making use of any (decent) starting guess at all \cite{Spence90,Spence98,Allen98,Spence99,Allen99, Allen00, Allen01_Systematic, Spence20,Spence21}. 

Approaches which use the sensitivity 
of multiple scattering to structure factor phases have proven 
useful also in X-ray diffraction \cite{Chang82,Juretschke82,Shen97}.  However, in contrast to Xray diffraction, the comparatively flat
Ewald sphere describing the scattering of high energy electrons usually causes more than just 3 reflections to be close to the Bragg condition, 
especially in complex structures, making these methods 
generally inapplicable
for phase determination from electron diffraction data.
Also in the field of low energy (reflection) electron diffraction (LEED) dynamical 
scattering is being applied for solving the phase problem \cite{Saldin02}.

The most significant contributions on the path to solving the dynamic inversion problem from transmission electron diffraction data have arguably been made by teams around Allen \cite{Allen98,Allen99, Allen00,Allen01_Systematic} and Spence \cite{Spence90,Spence98,Spence99,Spence20,Spence21}, inspiring each other and occasionally joining forces (on dynamic inversion from images rather than diffraction patterns) \cite{Allen01_Dyn}. Spence wrote of himself in a publication appearing in the last year of being among us, that he has worked on this subject for forty years of his life \cite{Spence21}. In addition to some early work on deriving expressions for gradients of the scattering matrix \cite{Spence90} he has contributed schemes that would exploit the interference in overlapping discs of CBED patterns \cite{Spence98}, or apply the principle of projection onto convex sets \cite{Spence99}. The last one of these approaches was picked up by him again very recently together with Donatelli, overcoming problems in convergence by exploiting the unitarity of matrix exponentials of hermition matrices \cite{Spence20,Spence21}. Under the assumption that both, absorption and the contribution of exciting additional diffracted beams to the diffraction intensities upon sequentially bringing all diffracted beams in a diffraction pattern into exact Bragg condition, can be neglected, they have been able to directly map a special tilt series of diffraction intensities to a unique set of structure factors. Again, inspiration from work by the group around Allen seemed important for this success \cite{Brown18}, who have also shown that absorption can be taken into account when making use of all the symmetries of the structure factor matrix \cite{Allen00}. 

The work presented  here is based on a joint publication of Spence and myself about a fast converging scattering path expansion of dynamical scattering \cite{Koch03,Koch02_dissertation}. Two different approaches, one based on directly solving the polynomial system of equations that result from this expansion, and a 'stacked scattering matrix' scheme based on a purely reciprocal space multislice formulation that can be solved by treating it like an artificial neural network, will be introduced. While the first approach requires the sample to be thin enough for kinematical scattering to still dominate the recorded intensities, the latter one is capable of dealing with much thicker samples. Demonstrations of these approaches will be based on simulated data in order to be able to compare to the ground truth in a more quantitative manner.

%%%%%%%%%%%%%%% FIG  %%%%%%%%%%%%%%%%%%%
\begin{figure}
%\begin{figure}[!tp]
%\ifnum \pictures = 1
\begin{center}
\includegraphics[width=0.78\textwidth]{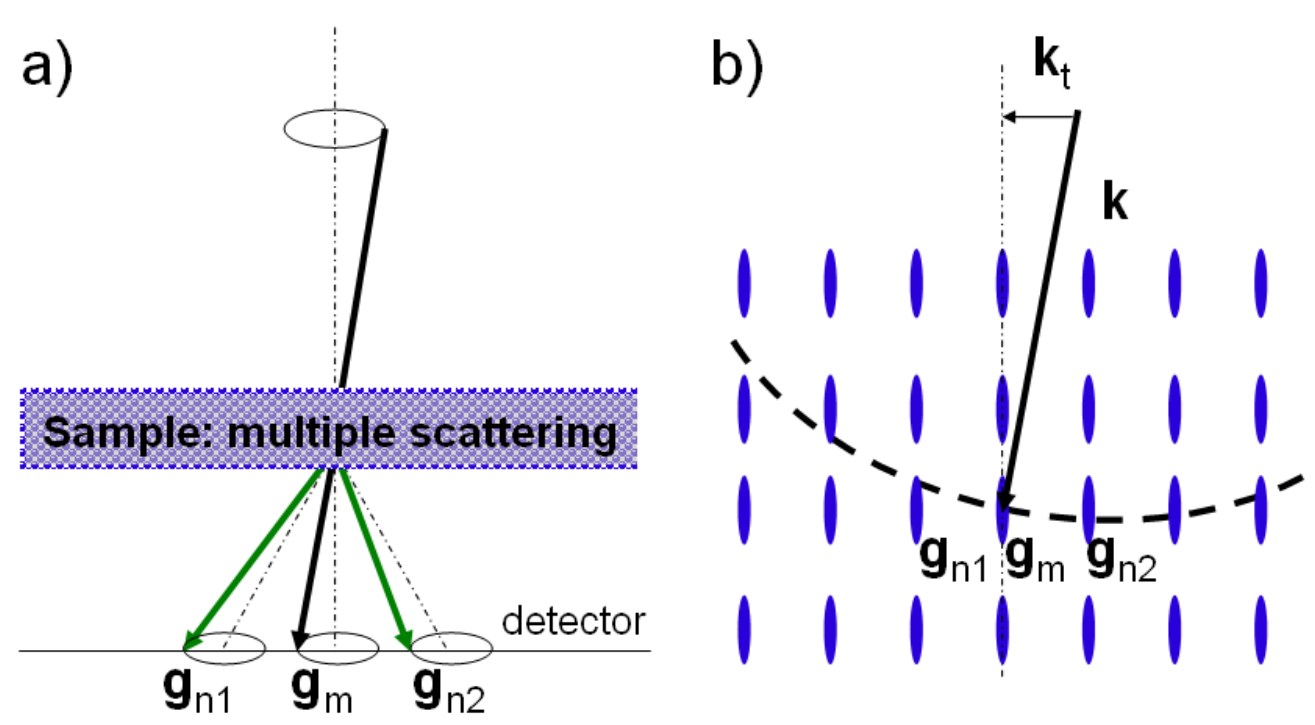}%
\end{center}
%\fi
\caption{Diagram illustrating the relevant parameters in real (a) and 
reciprocal space (b).
a) In convergent beam electron diffraction (CBED) experiments diffraction 
patterns for all incident beam directions whose 
\(\vec{k}_t\)-component lie within a disc are recorded. 
b) 
The reciprocal space representation illustrates how changing the incident 
beam direction causes the Ewald sphere (dashed arch) to intersect the 
crystal potential reciprocal lattice rods (relrods) with varying 
excitation errors (deviations from exact Bragg condition along the direction of the surface normal of the sample.} 
\label{geometry} \end{figure}

The experimental technique capable of acquiring the type of data that the following two dynamic inversion schemes require as input is large-angle rocking-beam electron diffraction (LARBED) \cite{Koch11}.
The advantage of LARBED patterns over conventional convergent-beam electron diffraction (CBED) patterns is that, although 
the radius of each individual diffraction disc may be several reciprocal 
lattice vectors (as in large-angle convergent-beam electron diffraction - LACBED), discs produced by the different diffraction spots do not 
overlap and may therefore be recorded in a single camera exposure.  This 
effect is produced by scanning the angle of
incidence on the sample of a nearly parallel beam during the exposure of the recording 
medium and partially de-scanning it again below the 
specimen.  
While the condition of avoiding overlapping discs makes CBED experiments nearly impossible if the unit cell size exceeds 1 nm, being able to control the reciprocal space range covered in LARBED data independent of the unit cell size makes this technique applicable to any material, including beam sensitive materials, since also the size of the illuminating spot may independently be adapted to the grain size of the specimen.
A reciprocal-space disc radius larger than half the distance between Bragg spots is also necessary to observe fluctuations in the 
diffraction intensities across the discs despite the low sample thickness 
required by the first inversion scheme to keep dynamical diffraction effects at a moderate level. The successful structure factor determination from experimental LARBED data by conjugate-gradient based minimization of the difference squared between the experimental data and a full Bloch wave simulation using 121 measured beams and 456 contributing structure factors has already been demonstrated for the simple centro-symmetric structure of SrTiO$_3$  and a fitted specimen thickness of 11.3 nm \cite{Wang16}. However, that approach was not yet successful in solving highly complex structures or retrieve structure factors from much thicker samples. 

\section{Expanding the scattering matrix}

Without loss of generality it is assumed that the specimen surface 
normal is parallel to the zone axis as well as the \(z\)-axis,
as illustrated in figure \ref{geometry}. 
The deviation of the 3-dimensional wave vector 
of
the incident electron beam \(\vec{k} = (k_x,k_y,k_z) = (\vec{k}_t, k_z)\) 
from the z-axis is then usually quite small.
Based on the Bloch wave solution to the Klein-Gordon equation of high 
energy scattering of a fast incident beam electron  by the crystal 
potential \cite{Fujimoto59,Humphreys79} the intensity
\(I_{g_n}  (\vec{k}_t)\)
of any point 
in the diffraction pattern
defined by the reciprocal lattice vector \(\vec{g}_n\) of the particular 
diffraction spot and the tangential component of the incident electron 
wave vector \(\vec{k}_t\) 
is given by the modulus squared of the element of the scattering matrix 
\begin{equation}
S(\vec{k}_t)_{n,m} = [e^{i T A}]_{n,m} 
\label{Eqn_Bloch}
\end{equation}
in the \(n^{th}\) row and the ``central beam'' column \(m\), i.e.\@ the beam 
index \(m\) for which the reciprocal lattice vector \(\vec{g}_m = 0\). 
Here \(A\) is a square matrix with the potential energy terms
\(A_{n,m}=U_{\vec{g}_n-\vec{g}_m} \) as its off-diagonal elements and 
the (relativistically corrected) terms 
\(\xi_n = -(|\vec{g}_n|^2+2\vec{g}_n \cdot \vec{k}) / \gamma \) 
related to the kinetic energy 
part of the modified Schroedinger equation along its 
diagonal.

The scalar parameter \(T= \pi \gamma \lambda t \) depends on 
the wavelength \(\lambda = (\lambda_0^{-2}+U_0)^{-1/2} \approx 
\lambda_0\)
of the incident electron beam corrected by the mean potential 
\(U_0\) of the crystal, the specimen thickness \(t\), and 
the relativistic correction factor \(\gamma=1+|e| v / m_0 c^2\) 
(here \(m_0\) and \(e\) are rest mass and charge of the  electron
and \(v\) is the accelerating voltage).  

Note that I did not multiply the electron structure factors 
\begin{equation}
U_{\vec{g}_n-\vec{g}_m}= 2 m_0 |e| V_{\vec{g}_n-\vec{g}_m} / h^2
\end{equation}
in \(A\)
with the usual relativistic correction factor \(\gamma\) but included it 
in \(T\) instead, in order to separate true
material constants (\(U_{\vec{g}}\)) from variable experimental parameters 
(\(\lambda(v), \gamma(v), t, \vec{k}_t\)).   
In the above expression 
\begin{equation}
V_{\vec{g}} = \Omega^{-1} \int_{cell} V(\vec{r}) \exp(2 \pi i \vec{g} \cdot \vec{r}) d^3\vec{r}
\end{equation}
are the Fourier coefficients of the crystal potential 
\(V(\vec{r})\), where the
integration is over one unit cell, and \(\Omega\) is the unit cell volume. 
The additive constant of \(U_0\) for every element along the diagonal of 
\(A\) has also been omitted because it only produces a general attenuation 
of the
diffraction pattern due to its imaginary part and a global phase 
offset due to the real part of \(U_0\), both unmeasurable in a 
conventional diffraction experiment. 

The crystal potential \(V(\vec{r}) = V^r(\vec{r})+i V^i(\vec{r})\) 
consists of a real part describing the elastic scattering and an imaginary 
part accounting for inelastic scattering processes which produce a 
non-isotropic attenuation in the scattered signal, if we collect the 
zero-loss scattered electrons only, as
is routinely done in today's quantitative (energy filtered) electron 
diffraction experiments.
This means that even for centro-symmetric crystals the \(U_{\vec{g}}\) are 
in general complex, and have a non-zero phase.

It has been shown that the Bloch wave expression (\ref{Eqn_Bloch}) can be 
expanded in a ''scattering path'' % or Born 
series \cite{Koch03}, 
providing a very general expression which agrees with those derived 
earlier by analyzing the multiple scattering process directly  
\cite{Cowley_Moodie,Fujiwara,Moodie72}.  Reference \cite{Koch03} 
provides a recursive, but finite explicit
expression for the coefficients \(C^{q}_{n,l_1,\ldots 
l_{q-1},m}(T,\vec{k}_t)\) in the expansion
% \begin{widetext}
\begin{eqnarray}  
|S(T,\vec{k}_t)_{n,m}|^2 =
|e^{T \xi_n(\vec{k}_t)}\delta_{n,m} + 
\sum_{q=1}^{\infty}
\sum_{l_1=0}^N \sum_{l_2=0}^N \cdots \sum_{l_{q-1}=0}^N \\
\nonumber 
\underbrace{U_{\vec{g}_n-\vec{g}_{l_1}}U_{\vec{g}_{l_1}-\vec{g}_{l_2}}
\cdots U_{\vec{g}_{l_{q-1}}-\vec{g}_m}}_{q} 
C^{q}_{n,l_1,\ldots l_{q-1},m}(T,\vec{k}_t)|^2
\label{BornExpansion}
\end{eqnarray}
% \end{widetext}
for arbitrary scattering path lengths, even in the (degenerate) case of 
multiple excitation of the same reflection.  The interested 
reader is referred to the original publication \cite{Koch03}, where 
the complete expression along with its derivation is provided.

\section{Inversion by polynomial equation solving}

The complex product of any two scattering path coefficients 
\begin{equation}
c^{q_1,q_2}_{n,l_1,\cdots,l_{q_1-1},h_1,\cdots,h_{q_2-1}}(T,\vec{k}_t)=C^{q_1}_{n,l_1,\ldots l_{q_1-1},m}(T,\vec{k}_t) \cdot C^{q_2}_{n,h_1,\ldots h_{q_2-1},m}(T,\vec{k}_t)^*
\end{equation}
defines the scale and phase factor with which 
any \(q_1+q_2=q\)-monomial of structure factors 
\(U_{\vec{g}_n-\vec{g}_{l_1}}
\cdots U_{\vec{g}_{l_{q_1-1}}-\vec{g}_m} 
U^*_{\vec{g}_n-\vec{g}_{h_1}}
\cdots U^*_{\vec{g}_{h_{q_2-1}}-\vec{g}_m}\) 
contributes to the 
diffraction intensity in the beam \(g_n\).

The convergence of above scattering path expansion (\ref{BornExpansion}) 
depends on                                                   
the value of \(T = \pi \gamma \lambda t\) and the modulus of the largest 
structure factor. Increasing the specimen thickness or decreasing the 
accelerating voltage will therefore increase the largest order of
significant monomials in the expansion.
We will let \(q_{max}\) be the length of the longest scattering path - or, 
in other words, the largest order of monomial - included in the expansion 
from here on. 
The coefficients \(c^{q_1,q_2} = c^{q_1,q_2}(T,\vec{k}_t)\) (\(q_1+ q_2 
\leq q_{max}\), I will omit
the functional dependence for increased readability from here on) depend 
in addition to the specimen thickness
and accelerating voltage also on the easily variable experimental 
parameters \(\vec{k}_t\).  Measuring diffraction intensities for a 
set of different values of incident electron beam tilt \(\vec{k}_t\), as 
in a CBED or LARBED experiment, allows a 
system of coupled multi-variate polynomial equations of degree
\(q_{max}\) to be defined whose solution is the set of structure factors 
which best (in a least squares sense) describe the observed intensities
for the set of experimental parameters.
The resulting system of equations can be solved by the 
reformulation-linearization technique
\cite{Sherali92} which has been proven to converge to a 
global optimum.

%%%%%%%%%%%%%%%%%%%%%%%%%%%%%%%%%%%%%%%%%%%%%%%%%%%%%
% C123 coefficients.

In order to keep the size of the set of polynomial equations within 
manageable limits, \(T\) should be small enough to let the first term in the 
expansion, the kinematic scattering intensity at reflection \(\vec{g}_n\), 
\(n \neq m\), \begin{eqnarray}
\nonumber 
I_{g_n}^{(2)}  (T,\vec{k}_t)
&=& \left| C^{1}_{n,m}(T,\vec{k}_t) U_{g_n}  \right|^2 =
\left| \frac{e^{T \xi_n} -1}{\xi_n} U_{g_n} \right|^2 \\ 
&=&    c^{1,1}_{n} |U_{g_n}|^2 
=
\frac{\gamma^2 \textrm{sin}^2(\pi  \lambda t s_{g_n})}{s_{g_n}^2}
|U_{g_n}|^2 
\label{eqn_kin_rocking_curve}
\end{eqnarray}
(\(s_{g_n} = \frac{1}{2} \xi_n / \gamma\) is also called the excitation 
error) %
be the dominant contribution, at least to the strong 
reflections. %
Although the kinematic scattering approximation \(I_{g_n}^{(q_{max}=2)}\)  
is not sensitive to any structure
factor phase invariants, it allows us to define sensible limits on the 
structure factor amplitude.

It also provides a very simple way of determining the specimen 
thickness, the only unknown parameter in the \(c^{q_1,q_2}\) coefficients 
(accelerating voltage and incident 
beam tilt can usually be controlled very precisely). 
By fitting the (normalized) experimental 2-dimensional rocking curves 
\(I_{g_n}(\vec{k}_t)\) for all
reflections with the kinematical one \(I_{g_n}^{(2)}  (T,\vec{k}_t)\) given by expression
(\ref{eqn_kin_rocking_curve}), both, the specimen thickness as well as a 
first estimate for the structure factor amplitude can be obtained \cite{Koch11}.  The 
form of expression (\ref{eqn_kin_rocking_curve})  allows this fit to be 
implemented very efficiently as a simple one-parameter search in 
the specimen
thickness \(t\).  The structure factor amplitudes are then determined 
directly by linear regression for each trial value of \(t\).    This is, by the way, a 
much
more precise measurement of structure factor amplitudes than using the 
integrated rocking curve as is done in precession electron diffraction 
\cite{Vincent94}.

% The \(q=2\) expansion coefficient 
% \begin{eqnarray}
% C^{2}_{n,l_1,m} &=& \left\{ \begin{array}{ll} 
% \frac{e^{T g_n}-1}{(g_n-g_{l_1})g_n} +
% \frac{e^{T g_{l_1}}-1}{(g_{l_1}-g_{n})g_{l_1}}
% & g_n \neq g_{l_1}\\
% e^{T g_n} \left(\frac{T}{g_n} - \frac{1}{g_n^2}\right) +\frac{1}{g_n^2}
% & g_n = g_{l_1} \\
% \end{array} \right. 
% \end{eqnarray}

Setting \(q_{max} = 3\) the scattering intensity at reflection 
\(\vec{g}_n\) (\(n \neq m\)) becomes
\begin{eqnarray}
\nonumber 
I_{g_n}^{(3)} (T,\vec{k}_t)
&=&   c^{1,1}_{n}
 |U_{g_n}|^2 + \sum_l  c^{2,1}_{n,l} 
U_{\vec{g}_n-\vec{g}_l}U_{\vec{g}_l}U_{\vec{g}_n}^* 
+ \\
\nonumber 
& & \sum_k  c^{1,2}_{n,h}
U_{\vec{g}_n}U_{\vec{g}_n-\vec{g}_h}^*U_{\vec{g}_h}^*  \\
\nonumber 
&=& I_{g_n}^{(2)} (T,\vec{k}_t)  + 2 \textrm{Re}
\left[ \sum_l  c^{2,1}_{n,l}
U_{\vec{g}_n-\vec{g}_l}U_{\vec{g}_l}U_{\vec{g}_n}^* \right]
\end{eqnarray}
The monomials \(U_{\vec{g}_n-\vec{g}_l}U_{\vec{g}_l}U_{\vec{g}_n}^*\) can 
be written as the product of the 3 structure factor amplitudes and a phase 
factor whose argument is the 3-phase invariant 
\(\phi_{\vec{g}_n-\vec{g}_l} + \phi_{\vec{g}_l} - \phi_{\vec{g}_n}\), 
indicating that  \(I_{g_n}^{(3)} (T,\vec{k}_t)\) must also be sensitive to 
these phase invariants.

%%%%%%%%%%%%%%%%%%%%%%%%%%%%%%%%%%%%%%%%%%%%%%%%%%%%%
% Solving the polynomial equations

The reformulation-linearization technique developed by Sherali and 
Tuncbilek \cite{Sherali92} converts the non-linear polynomial equations 
into linear ones by treating each monomial as an independent variable.  By 
defining additional linear equations that result from obvious 
relationships between the monomials as well as inequalities produced by 
upper and lower bounds on the variables, the polynomial set of equations is 
converted into a set of linear equalities and inequalities for which a 
solution can be found, if one exists, using standard linear programming 
algorithms.  Once a solution has been found, the limits on the variables 
may be readjusted (''branch and bound''), resulting in a new linear 
programming problem and eventually the global solution.

If the initial estimate of the structure factor amplitudes is fairly 
reliable, the limits within which the correct structure factor 
amplitude must lie may be defined rather tightly.  It is then very likely 
that the correct solution will be found already in the first 
linearization step, as has been the case for the test case described 
below.

% Koch05_2dcryst
%%%%%%%%%%%%%%%%%%%%%%%%%%%%%%%%%%%%%%%%%%%%%%%%%%%%%
% Test on simulated data

In order to verify the performance of the phase retrieval method described 
above it has been tested using simulated electron diffraction data.
Since it has already been shown 
that Friedel asymmetries in dynamical diffraction patterns 
of non-centrosymmetric structures can be used to solve the phase 
problem in the framework of this expansion \cite{Koch05_2dcryst}, I used simulated diffraction patterns of Si in 
the (110) projection, a simple centro-symmetric structure for 
demonstration purposes.
If absorption is neglected, as has been done for this test case, kinematic 
and higher order scattering contributions to the diffraction intensities 
cannot be separated simply on the basis of their symmetry, as it is 
partly the case
for non-centro-symmetric structures \cite{Koch05_2dcryst}, making the 
reconstruction of centro-symmetric structures even more challenging.
In the remainder of this section I demonstrate the direct retrieval of the structure factors of silicon from simulated LARBED data.

%%%%%%%%%%%%%%% FIG  %%%%%%%%%%%%%%%%%%%
\begin{figure}
%\begin{figure}[!tp]
%\ifnum \pictures = 1
\begin{center}
\includegraphics[width=0.78\textwidth]{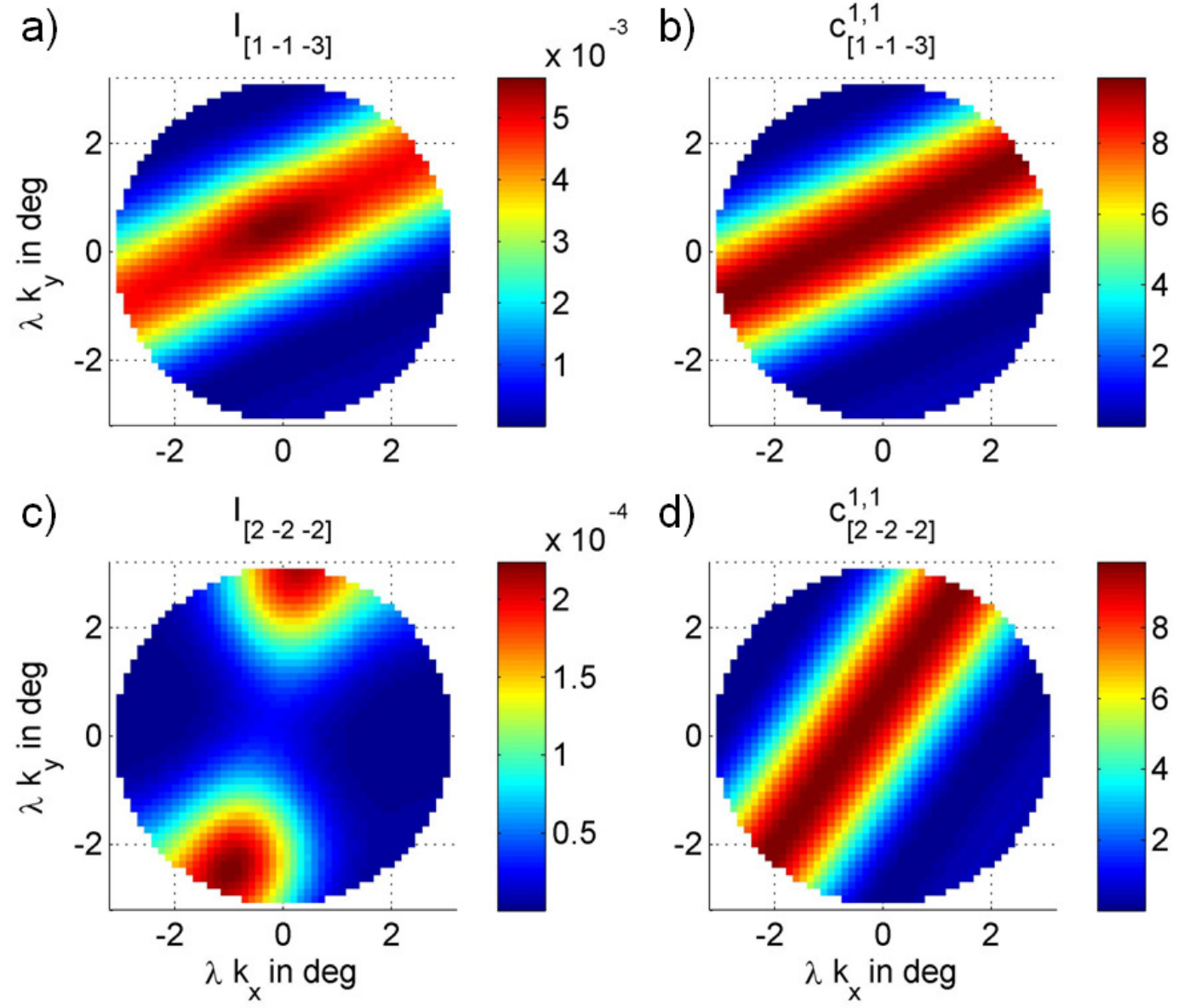}%
\end{center}
%\fi
\caption{a) Simulated LARBED disc for \(\vec{g}=[1 \bar{1} \bar{3}]\). 
b) The coefficient \(c^{1,1}_{[1 \bar{1} \bar{3}]}(t=3.985 \textrm{nm},\vec{k}_t)\) given by 
expression \ref{eqn_kin_rocking_curve} (unscaled kinematic rocking curve).  
The
square root of the proportionality constant between the 2 plots shown in 
a) and b) provides a first estimate of the structure factor amplitude 
for this reflection.
c) Simulated LARBED disc for the kinematically forbidden reflection \([2 
\bar{2} \bar{2}]\). Comparing the diffraction intensity of a forbidden 
reflections with the corresponding kinematic rocking curve (d) allows 
its identification as being forbidden to be automated.} 
\label{discs} \end{figure}

LARBED disc intensities for the following 27 beams have been computed by 
Bloch wave simulation: 
\([\bar{2}   2   2]\),
\([\bar{2}   2  \bar{2}]\),
\([ 0   0  \bar{2}]\),
\([\bar{3}   3   1]\),
\([\bar{3}   3  \bar{1}]\),
\([\bar{2}   2   4]\),
\([\bar{2}   2  \bar{4}]\),
\([\bar{1}   1   3]\),
\([\bar{1}   1  \bar{3}]\),
\([ 0   0  \bar{4}]\),
\([\bar{2}   2   0]\),
\([\bar{1}   1   1]\),
\([\bar{1}   1  \bar{1}]\),
\([ 0   0   0]\),
\([ 1  \bar{1}   1]\),
\([ 1  \bar{1}  \bar{1}]\),
 \([2  \bar{2}   0]\),
 \([0   0   4]\),
 \([1  \bar{1}   3]\),
 \([1  \bar{1}  \bar{3}]\),
 \([2  \bar{2}   4]\),
 \([2  \bar{2}  \bar{4}]\),
 \([3  \bar{3}   1]\),
 \([3  \bar{3}  \bar{1}]\),
 \([0   0   2]\),
 \([2  \bar{2}   2]\),
 \([2  \bar{2}  \bar{2}]\).          
An accelerating voltage of 200 kV, a specimen thickness of 4 nm and a disc 
radius of \(3^\circ\)
have been used, a beam tilt angle which is easily achievable in actual 
experiments with the acquisition method described in the previous 
paragraph (LARBED disc radii larger than \(3^\circ\) have been demonstrated experimentally \cite{Koch11}). %

Examples of the resulting 
diffraction data are shown in figure \ref{discs} a and c.  
While the high symmetry of centro-symmetric structures removes 
potentially useful Friedel asymmetries in the diffraction data, it also 
imposes symmetry constraints on the structure factors themselves.  
The
26 structure factors corresponding to the reflections listed above are
reduced to 8 independent real-valued variables (\(U_0\) was set to 0 for 
reasons mentioned earlier).

In step 1 of the reconstruction the specimen thickness and initial 
estimates of the structure factor amplitudes were obtained by fitting 
kinematic rocking curves (see figure \ref{discs} b and d) to the simulated 
LARBED discs by nonlinear single parameter least-squares fitting 
w.r.t. the specimen thickness as described earlier, automatically 
removing forbidden reflections from the fit based on their poor agreement 
with the shape of the kinematic rocking curves. 
The best matching thickness was found to be 3.985 nm, deviating only 0.3 
\% 
from the thickness used for the simulation, most likely due to 
dynamical effects.

In step 2 of the reconstruction the previously determined sample thickness 
of 3.985 nm was used to define the \(c^{q1,q2}_{g_n}(T,\vec{k}_t)\) 
coefficients.
Both, the definition of these 
coefficients as well as the setup of the linearized system of polynomial 
equations is fully automated, requiring as only input the 
specimen thickness, the reciprocal 
lattice vectors to be included in the reconstruction, the deviation from 
zone axis \(\vec{k}_t\)
for each point within the LARBED discs and the 
diffraction intensity at that point.
All of these parameters can be obtained 
directly from the (calibrated) diffraction pattern.

%%%%%%%%%%%%%%% FIG  %%%%%%%%%%%%%%%%%%%
\begin{figure}[t]
%\begin{figure}[!tp]
%\ifnum \pictures = 1
\begin{center}
\includegraphics[width=0.78\textwidth]{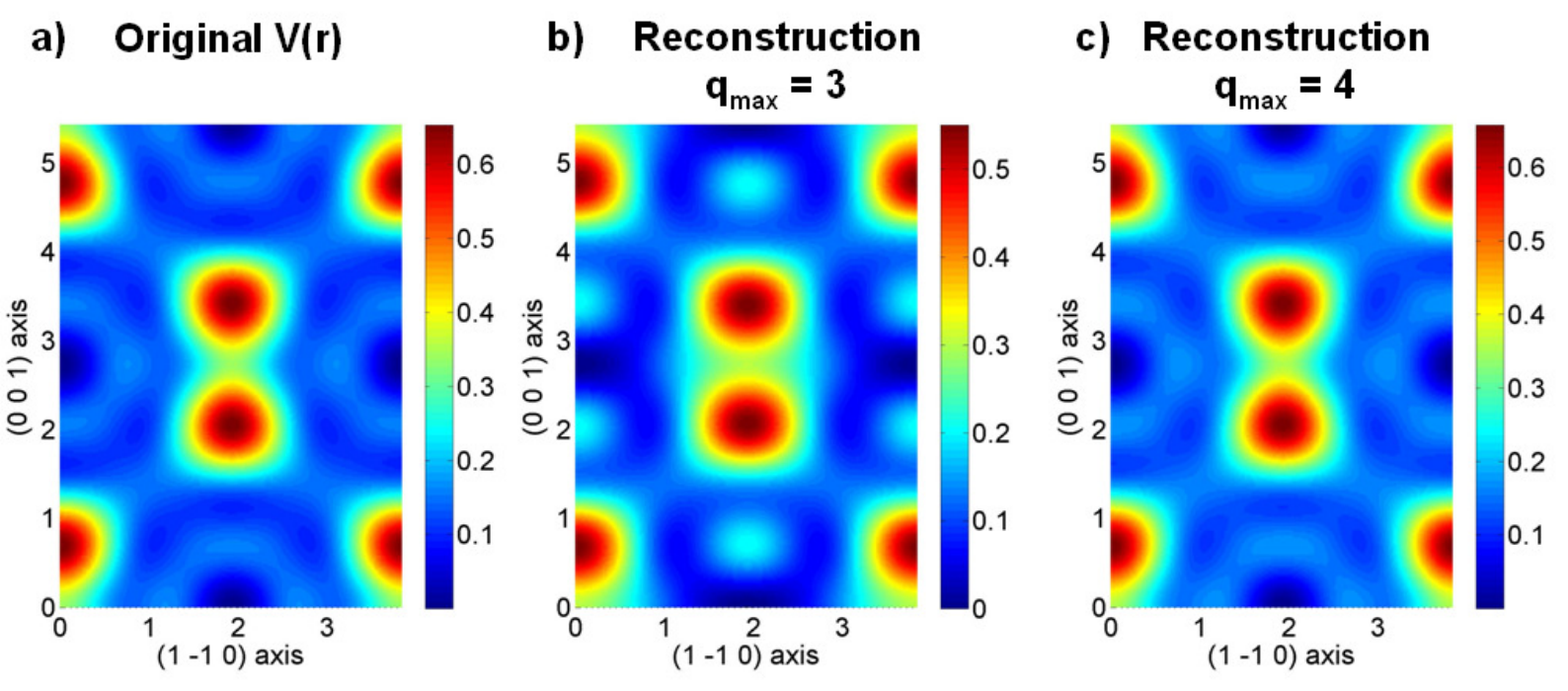}%
\end{center}
%\fi
\caption{Reconstructed real-space potential distribution within a single 
unit cell of Si in the (110) projection. a) Potential distribution 
used to simulate the diffraction data.  b) Reconstruction from a 
scattering path expansion with \(q_{max}=3\), i.e.\@ one scattering order 
beyond the kinematic approximation.  It is quite obvious that this order 
of approximation is not sufficient. c) Reconstruction from a 
scattering path expansion with \(q_{max}=4\) showing excellent 
agreement with the original.  Since the value of
\(U_0\) cannot be reconstructed from the available data, the potential is 
displayed with its minimum value set to 0 in all 3 figures.} 
\label{reconstruction} 
\end{figure}

In step 3 of the reconstruction the previously defined system of linear 
equations and inequalities was solved using the lsqlin function provided by the Matlab optimization 
toolbox, which applies an active set method for solving linear programs in 
its medium-scale version.  Repeating the reconstruction for \(q_{max}=3\) 
and \(q_{max}=4\) resulted in the projected potential plots shown in 
Figs. \ref{reconstruction}b and \ref{reconstruction}c respectively.  
The 
reconstruction for \(q_{max}=4\) involved 33 different monomials, 11670 
equalities and 122 inequalities and took about 1 second on a 2.2 GHz 
laptop equipped with 500 MB RAM to compute, solving for magnitude and 
sign of 6 distinct structure factors (being equal to zero, the structure 
factors of the
forbidden reflections had been removed automatically from the fit in step 
1 of the reconstruction).

The agreement of the second reconstruction including scattering paths up 
to length 4 with the original potential is excellent, demonstrating that 
this method works very well, provided that scattering paths of 
sufficiently high order are included.
Increasing the number of reflections increases the number of unknowns, 
but also the number of available data.  However, the number of 
monomials,
equalities and inequalities increases as the \(q_{max}^{th}\) 
power
of the number of structure factors to be determined.  It will therefore 
be necessary to define a scheme by which the dominant monomials and the 
linear relationships
between them will be kept, discarding those which are not important. It thus does not seem very practical to go to orders of \(q_{max}\) greater than 4 or 5 and thicknesses where scattering orders up to \(q_{max}\) dominate. For thicker crystals, one can reformulate the description of a single crystalline slab into a stack of many equal crystalline slabs, as is the basis for the approach described in the following section.

\section{Inversion by the stacked Bloch wave approach}

Based on the multislice approach to solving the Schr\"odinger equation describing the scattering of the fast electron in the crystal potential, an inversion scheme for retrieving the 3-dimensional object potential from a tilt series (tilt limited to $\pm 10^{\circ}$) of high-resolution TEM (HRTEM) images has been demonstrated on simulated data \cite{vandenbroek2012,VandenBroek13}. The experimental realization of this approach has so far remained elusive, very likely due to the difficulty to align HRTEM images acquired at different tilts with a sufficiently high precision. Fortunately, the problem of aligning images does not occur in the case of LARBED data. 

While the standard multislice approach is based on a real-space description of the scattering potential corresponding to the tilted crystal and the propagation between slices using the convolution of the scattered wave function by the Fresnel propagator, we may also describe the scattering within each of the slices in reciprocal space, using the Bloch wave formalism, resulting in a 'stacked Bloch wave' formulation \cite{Pennington14} with the scattering matrix of the crystal given by
\begin{equation}
S(\vec{k}_t) = [e^{i T A(\vec{k}_t)}] = \prod_{j=1}^{N_{slices}} S^{(j)}(\vec{k}_t) = \prod_{j=1}^{N_{slices}} [e^{i T A^{(j)}(\vec{k}_t)}]
\end{equation}   

The structure factor matrices $A^{(j)}(\vec{k}_t)$ describing each of the slices in this formulation do not have to be the same. In the presence of just a few layers, a sufficient range of tilt angles, and data with a low level of noise, it is possible to retrieve parameters affecting the entries in the individual $A^{(j)}$-matrices, including the correct sequence \cite{Pennington15a,Pennington15,Pennington18}. However, in the context of solving the structure of a single crystal it makes sense to require that all $N_{slices}$ structure factor matrices are the same. This also allows us to make $N_{slices}$ very large ($N_{slices}$ was set to 45 in the example below). 

We can now chose any scattering path length for approximating the individual scattering matrices $S^{(j)}(\vec{k}_t)$. If $N_{slices} > 1$ we can also use a scattering path length of one in each layer, since, in contrast to expression (\ref{eqn_kin_rocking_curve}), which corresponds to $N_{slices} = 1$ the sensitivity to the structure factor phases is brought in by the fact that multiple scattering is also accounted for by having multiple slices. One advantage of using a single scattering approximation 
\begin{equation}
S^{(j)}(\vec{k}_t)_{n,m} \approx \left\{
\begin{array}{ c l }
1 & \textrm{if } n=m\\
\frac{e^{T \xi_n} -e^{T \xi_m}}{\xi_n-\xi_m} U^{(j)}_{g_n-g_m} & \textrm{if } \xi_n \neq \xi_n \\
i T e^{T \xi_n} U^{(j)}_{g_n-g_m} & \textrm{if } \xi_n = \xi_m
\end{array}
\right.
\label{eqn:FOEBW}
\end{equation}
for each layer is also that the analytical expression for the gradient of the scattering matrix w.r.t. the individual structure factors is very straight forward:    
\begin{equation}
\frac{\partial S^{(j)}(\vec{k}_t)_{n,m}}{\partial U^{(j)}_{g_n-g_m}} \approx \left\{
\begin{array}{ c l }
0 & \textrm{if } n=m\\
\frac{e^{T \xi_n} -e^{T \xi_m}}{\xi_n-\xi_m}  & \textrm{if } \xi_n \neq \xi_n \\
i T e^{T \xi_n}  & \textrm{if } \xi_n = \xi_m
\end{array}
\right.
\end{equation}
Having the above expressions for $S^{(j)}(\vec{k}_t)$ and $\frac{\partial S^{(j)}(\vec{k}_t)_{n,m}}{\partial U^{(j)}_{g_n-g_m}}$ we can follow the procedure minimizing the discrepancy between the experimental LARBED intensities and the intensities being simulated using the stacked single scattering path approximation using any gradient descent method (e.g.\@ iRProp+ \cite{iRProp}, as in the case described below).

Note that, in order to keep the analytical expression for $\frac{\partial S^{(j)}(\vec{k}_t)_{n,m}}{\partial U^{(j)}_{g_n-g_m}}$ each layer has to have its own set of structure factors $U^{(j)}_{g_n-g_m}$. For a limited tilt range and a large number of layers this would in most cases result in over-fitting, if corresponding structure factors in different layers are not constrained to be the same. This can be done by including an explicit constraint in the error metric being minimized, or by replacing the current estimate of the structure factors by their average every few iterations of the optimization procedure.    

%%%%%%%%%%%%%% FIG  %%%%%%%%%%%%%%%%%%%
\begin{figure}
%\begin{figure}[!tp]
%\ifnum \pictures = 1
\begin{center}
\includegraphics[width=0.9\textwidth]{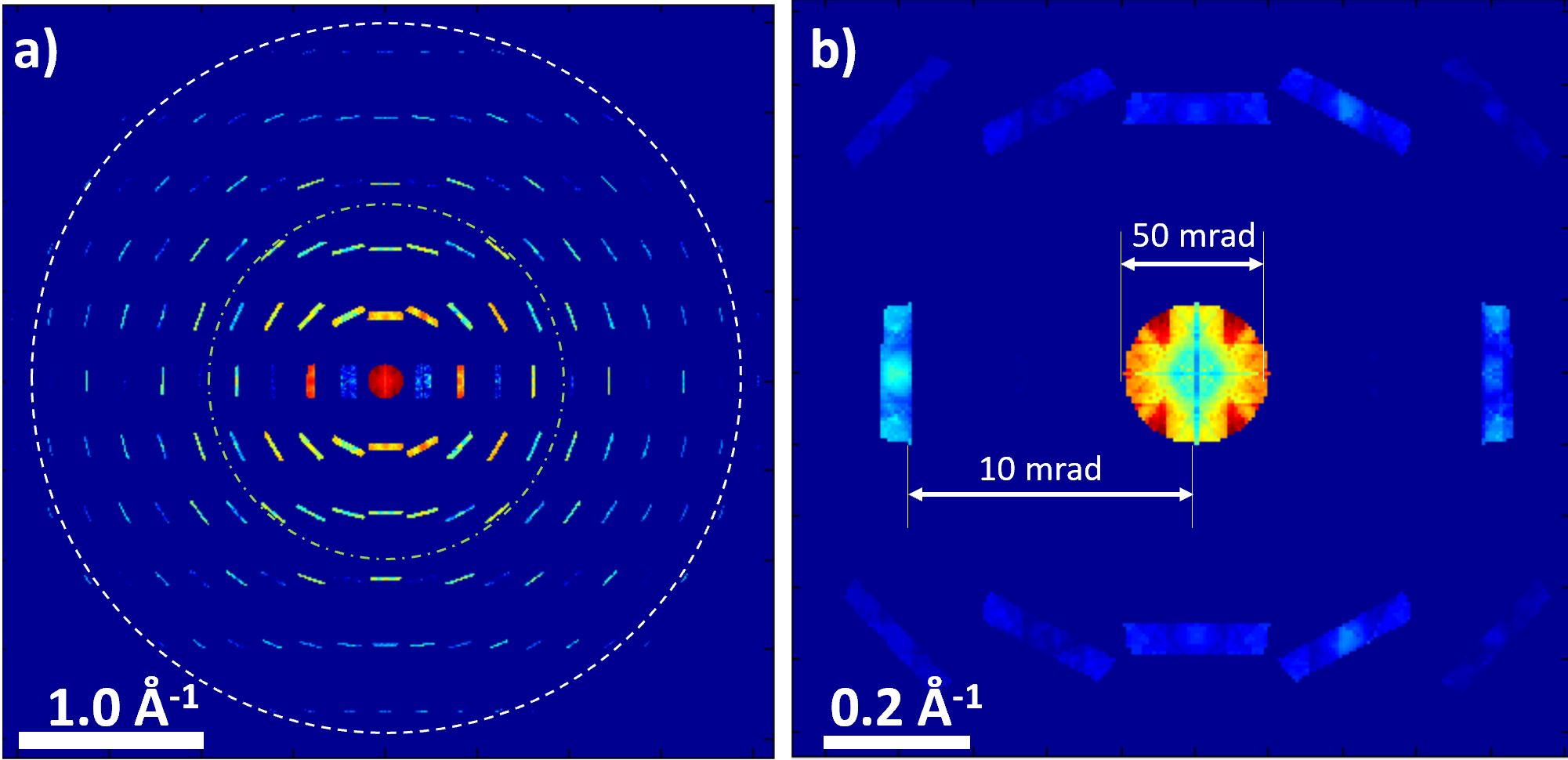}%
\end{center}
%\fi
\caption{Simulated LARBED pattern of GaN (110) (simulation parameters in text). a) Complete diffraction pattern on a logarithmic scale: The largest g-vectors included as diffracted beams in this Bloch wave simulation have a reciprocal space length of 2\AA$^{-1}$, as indicated by the white dashed circle. The green dash-dotted circle indicates the tilt range of the illumination during the LARBED simulation. b) Central part of the diffraction pattern on a linear scale: The geometry of the representation of the LARBD dataset is illustrated by highlighting that the scale  of the LARBED discs is 10 times larger than that of the distance between discs (descan = 0.9 \cite{Koch11}).  } 
\label{FIG:GaN_DP} \end{figure}

Fig. \ref{FIG:GaN_DP} shows a simulated LARBED pattern of 10 nm thick GaN in the (110) zone axis. The accelerating voltage for this simulation was 200 kV and the maximum g-vector of reflections being simulated was 2\AA$^{-1}$. The Bloch wave method was used for this simulation, where structure factors on reciprocal lattice points up to 4\AA$^{-1}$ from the origin were included in the structure factor matrix. A total of 194 different structure factors (not counting $U_0$) were included in the calculation. The selection of beams that contributed to the diffraction pattern was carried out separately for each beam tilt, including only beams that were closer than 0.002\AA$^{-1}$ to the Ewald sphere. This kept the number of beams in the simulation for each beam tilt relatively low but also leads to the sharp cutoff of diffraction discs, causing them to show mostly the Bragg lines (line in reciprocal space, along which the Bragg condition is very close to being matched for that reflection).

%%%%%%%%%%%%%% FIG  %%%%%%%%%%%%%%%%%%%
\begin{figure}
%\begin{figure}[!tp]
%\ifnum \pictures = 1
\begin{center}
\includegraphics[width=0.9\textwidth]{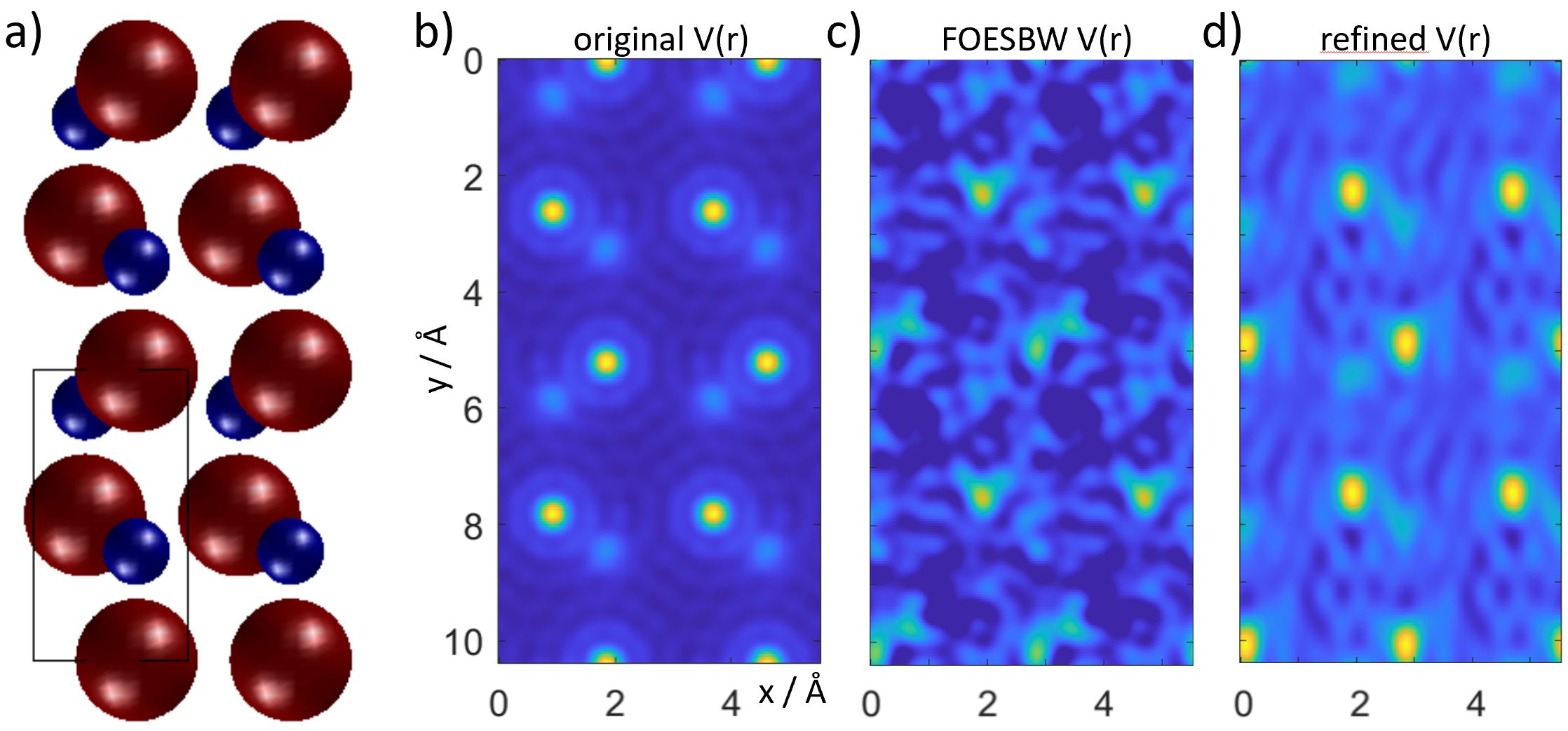}%
\end{center}
%\fi
\caption{Reconstruction of the crystal potential from a simulated LARBED dataset: a) Structure of GaN in the (110) zone axis. 2 x 2 unit cells are being displayed, as is the case for Figs. b) - d). b) Original crystal potential simulated using independent atom scattering factors. c) Crystal potential resulting from all 194 structure factors that have been reconstructed using the first-order expansion stacked Bloch wave (FOESBW) approach described in this work. d) Bandwidth-limited potential resulting from a set of structure factors that was refined from the FOESBW-reconstruction using a simplex optimization of the full scattering matrix. The potential has been bandwidth limited to 2\AA$^{-1}$, i.e.\@ up to where reflection spots were simulated. All three potential maps are displayed using the same color map and -range.} 
\label{FIG:GaN_Rec} \end{figure}

Using the first order expansion of the Bloch wave formalism (eqn. (\ref{eqn:FOEBW})) in each layer, 45 layers were stacked, each having its independent set of structure factors. Having also analytical expressions for the gradient, gradient-based optimization was applied to find the structure factors in each of the layers. A constraint penalizing differences of corresponding structure factors in neighboring layers was added, initially with a very low weight, but increasing in weight, every time the cost function went below a given threshold. Since the diagonal entries of the first-order approximated scattering matrices given by eqn. (\ref{eqn:FOEBW}) are actually zeroth order in structure factors (the 1st order vanishes), these values were also optimized for, even though they do not depend on any structure factor. Every 250 iterations corresponding elements of these diagonal entries were set to their average value.

Starting from a set of 194 structure factors which consisted of amplitudes fitted to the LARBED data using expression (\ref{eqn_kin_rocking_curve}) and fully random structure factor phases, Fig. \ref{FIG:GaN_Rec}c shows a real-space representation of the average potential resulting from the first-order expansion stacked Bloch wave (FOESBW) optimization. Since the approximation to a full Bloch wave calculation is rather poor, the reconstruction is not expected to yield perfect agreement. However, in addition to the position of the Ga atoms, also the rough positions of the N atoms are visible, despite their more than 4 times lower Z-number. This set of complex-valued structure factors is then used as starting guess of a Nelder-Mead Simplex optimization using the Matlab function fminsearch. A bandwidth-limited version of the refined potential is shown in Fig. \ref{FIG:GaN_Rec}d. The bandwidth limiting seemed appropriate, because those structure factors that did not correspond to any diffracted intensity, but contributed only indirectly were only weakly determined due to the relatively low thickness.

Apart from a shift in unit cell origin which cannot be determined from the diffraction data, the agreement of the refined potential and the original one used as input for generating the simulated data is rather good. Note that the structure factors were not constrained in any way, neither by enforcing a real-valued (purely elastic) or positive potential, nor by assuming peakedness (atomicity). Since the imaginary part of the reconstructed potential consists of small numbers around zero, it is not shown here. By not assuming any atomicity in the reconstruction, this method is also applicable to highly beam-sensitive materials, where it may not be possible to collect diffraction data at sufficiently high resolution for atomicity-based phasing techniques to become applicable.

%%%%%%%%%%%%%%%%%%%%%%%%%%%%%%%%%%%%%%%%%%%%%%%%%%%%%
\section{Conclusion}

In summary, two new methods for solving the inversion problem of dynamical 
 scattering from LARBED data have been presented. 
Apart from the usually well-defined accelerating voltage the data required 
by
the reconstruction may be directly obtained from the diffraction pattern alone.  
No assumptions are being made about the scattering potential 
within a single unit cell, making this method much more general than 
direct phasing methods applied to kinematical diffraction data, where a high degree of completeness and sufficient resolution for applying atomicity constraints are being required.
The method uses dynamical scattering effects in the diffraction data to 
solve directly for the structure factor amplitudes and phases.

Although the application of this method to X-ray diffraction patterns is 
quite conceivable, it is expected that it's main application will be in 
the field of structural electron crystallography.  Simulated 
LARBED data of Si (110) and GaN (110) have been  used to verify these methods.

\bibliography{TEM}

\end{document}